\documentclass[twocolumn,trackchanges]{aastex6}

\usepackage{epsfig,graphicx,amssymb,epstopdf,mathrsfs,float,url}
\newcommand{\enu}{\varepsilon_\nu}
\newcommand{\egam}{\varepsilon_\gamma}

\newcommand{\fermi}{{\em Fermi}}

\newcommand{\ice}{IceCube}
\newcommand{\antares}{ANTARES}
\newcommand{\icfnorth}{\mbox{IC59-North}}
\newcommand{\icfsouth}{\mbox{IC59-South}}
\newcommand{\txsblazar}{TXS~0506+056}

\newcommand{\adstat}{Anderson-Darling}
\newcommand{\nugamma}{\mbox{$\nu$+$\gamma$}}
\newcommand{\lamten}{\mbox{$\lambda_{\times 10}$}}
\newcommand{\lamhund}{\mbox{$\lambda_{\times 100}$}}
\newcommand{\ngexp}{\mbox{$\langle n_\gamma \rangle$}}
\newcommand{\ninj}{\mbox{$n_{\rm inj}$}}
\newcommand{\nobs}{\mbox{$n_{\rm obs}$}}

\def\arcdeg{\hbox{$^\circ$}}

\def\simlt{\mathrel{\hbox{\rlap{\hbox{\lower4pt\hbox{$\sim$}}}\hbox{$<$}}}}
\def\simgt{\mathrel{\hbox{\rlap{\hbox{\lower4pt\hbox{$\sim$}}}\hbox{$>$}}}}

\begin{document}

\title{A Coincidence Search for Cosmic Neutrino and Gamma-Ray Emitting 
  Sources using IceCube and Fermi LAT Public Data}

\email{cft114@psu.edu}

\author{C.~F. Turley\altaffilmark{1,3},
        D.~B. Fox\altaffilmark{2,3,4},
        A. Keivani\altaffilmark{1,3},
        J.~J. DeLaunay\altaffilmark{1,3},
        D.~F. Cowen\altaffilmark{1,2,3},
        M. Mostaf\'{a}\altaffilmark{1,2,3}, \\
        H.~A. Ayala Solares\altaffilmark{1,3},
        \& K. Murase\altaffilmark{1,2,3}
}


\affil{$^1$Department of Physics, Pennsylvania State University,
           University Park, PA 16802, USA \\
       $^2$Department of Astronomy \& Astrophysics, Pennsylvania
           State University, University Park, PA 16802, USA \\
       $^3$Center for Particle \& Gravitational Astrophysics,
           Institute for Gravitation and the Cosmos, Pennsylvania
           State University, University Park, PA 16802, USA \\
       $^4$Center for Theoretical \& Observational Cosmology,
           Institute for Gravitation and the Cosmos, Pennsylvania
           State University, University Park, PA 16802, USA}

%


\begin{abstract}

  We present results of an archival coincidence analysis between
  \fermi\ Large Area Telescope (LAT) gamma-ray data and public
  neutrino data from the \ice\ neutrino observatory's 40-string (IC40)
  and 59-string (IC59) observing runs. Our analysis has the potential
  to detect either a statistical excess of neutrino + gamma-ray
  (\nugamma) emitting transients or, alternatively, individual high
  gamma-multiplicity events, as might be produced by a neutrino
  observed by \ice\ coinciding with a LAT-detected gamma-ray
  burst. Dividing the neutrino data into three datasets by hemisphere
  (IC40, \icfnorth, and \icfsouth), we construct uncorrelated null
  distributions by Monte Carlo scrambling of the neutrino datasets. We
  carry out signal-injection studies against these null distributions,
  demonstrating sensitivity to individual \nugamma\ events of
  sufficient gamma-ray multiplicity, and to \nugamma\ transient
  populations responsible for $>$13\% (IC40), $>$9\% (\icfnorth), or
  $>$8\% (\icfsouth) of the gamma-coincident neutrinos observed in
  these datasets, respectively. Analyzing the unscrambled neutrino
  data, we identify no individual high-significance neutrino + high
  gamma-multiplicity events, and no significant deviations from the
  test statistic null distributions. However, we observe a similar and
  unexpected pattern in the \icfnorth\ and \icfsouth\ residual
  distributions that we conclude reflects a possible correlation
  ($p=7.0\%$) between IC59 neutrino positions and persistently bright
  portions of the \fermi\ gamma-ray sky. This possible correlation
  should be readily testable using eight years of further data already
  collected by \ice. We are currently working with Astrophysical
  Multimessenger Observatory Network (AMON) partner facilities to
  generate low-latency \nugamma\ alerts from \fermi\ LAT gamma-ray and
  \ice\ and \antares\ neutrino data and distribute these in real time
  to AMON follow-up partners.
   
\end{abstract}

\keywords{BL Lacertae objects: general --- %
          cosmic rays --- %
          gamma-rays: bursts --- %
          gamma-rays: general --- %
          neutrinos} 

\maketitle


\section{Introduction}
\label{sec:intro}

The \ice\ Collaboration has detected the first high-energy neutrinos
of cosmic origin \citep{ic-pev,Ic3+13-sci}. Unlike the atmospheric
neutrinos that dominate the observed events at lower energies, the
cosmic neutrinos have a harder spectrum, with a current best-fit
neutrino power-law index of $\Gamma_\nu = -2.19$
\citep{icecubeicrc17}. The sky distribution of high-likelihood cosmic
neutrinos is consistent with isotropy, and indeed, no high-confidence
counterparts have been identified for any of these neutrinos
\citep{icpoint2017,icfouryears}; however, we note the recent
suggestive coincidence between the ``Extremely High Energy'' muon
neutrino \mbox{IceCube-170922A} \citep{170922gcn} and a bright and
extended GeV-flaring episode of the blazar
\txsblazar\ \citep{1709blazaratel}.

In addition to blazars, possible cosmic neutrino source populations
include star-forming and starburst galaxies, galaxy groups and
clusters, other types of active galactic nuclei, supernovae, and
gamma-ray bursts (see \citealt{muraseorigin} for a recent review). 

One possible approach to revealing the nature of the source
population(s) is to take advantage of the likely-greater number of
cosmic neutrinos that must exist within the \ice\ dataset at lower
energies. For example, using the most recent power-law index and
normalization estimates \citep{icecubeicrc17}, and integrating down to
$\enu \approx 1$\,TeV using the facility's declination- and
energy-dependent effective area \citep{icfouryears}, we find that
\ice\ should be detecting $r_{\rm cosmic} \approx 120$ neutrinos of
cosmic origin per year, all-sky, below the $\enu \approx 60$\,TeV
threshold for individual likely-cosmic events (e.g., those selected as
\ice\ High Energy Starting Events). If the cosmic neutrino spectrum
softens (or becomes dominated by a distinct, softer component) within
${\rm 1\,TeV} \simlt \enu \simlt {\rm 60\,TeV}$ then the number of
cosmic neutrinos in this range could be substantially
greater. However, since these lower-energy cosmic neutrinos are
individually indistinguishable from the atmospheric neutrino
background, some strategy must be employed to identify them before
they can be used to study their sources.

One such strategy is illustrated by the \ice\ Collaboration's all-sky
and catalog-based point-source searches
\citep{galacticnu2017,icpoint2017,icfouryears}. These strategies are
likely to be optimal in cases where the neutrino sources are
persistent, roughly constant, and drawn (respectively) from either
unknown or known/anticipated source populations.

Alternatively, for transient or highly-variable source populations, we
can take advantage of the neutrino timing and localization to attempt
to identify electromagnetic or other non-neutrino counterparts. Any
such discovery would have immediate implications for the nature of the
sources, whether or not a host galaxy or long-lived counterpart could 
also be identified. 

As reviewed by \citet{muraseorigin}, numerous theoretical models
predict the co-production of cosmic neutrinos with prompt and luminous
electromagnetic signals. In most such models, protons or other nuclei
are accelerated to high energies, often in relativistic
jets. Interactions of these accelerated particles with ambient matter
or radiation yield copious quantities of pions, with gamma-rays
resulting from decay of the $\pi^0$ component, and neutrinos from
decays of the co-produced $\pi^\pm$. For example, gamma-ray bursts
(GRBs) were long considered potential sources of jointly-detected
high-energy neutrinos and gamma-rays (e.g.,
\citealt{wbfireball,bbm+15,meszaros15}). Although GRBs are now ruled
out as the dominant source of \ice\ cosmic neutrinos by coincidence
studies \citep{icgrblimits}, it remains possible that GRBs supply a
fraction of the cosmic neutrinos.

Particular sub-classes of GRBs including ``choked jet'' events could
still provide a partial or even dominant contribution to the cosmic
neutrinos \citep{muraseioka13,smm16,tamborraando16}.  Other promising
neutrino + gamma-ray (\nugamma) transients include luminous supernovae
\citep{murase+11}, blazar flares (e.g., \citealt{dmi14,gpw17}), and
tidal disruption events (e.g.,
\citealt{daifang17,smm17,lunardiniwinter17}).

In this context, the \fermi\ satellite's Large Area Telescope (LAT;
\citealt{latdesg}) offers a highly complementary dataset for
cross-reference with \ice\ neutrino detections. Operating efficiently
over the ${\rm 100\,MeV} \simlt \egam \simlt {\rm 300\,GeV}$ energy
range, the LAT provides instantaneous coverage of roughly 20\% of the
sky and regular full-sky coverage (under normal operations) every
three hours. Its energy range, angular and energy resolution, low
background, and sensitivity yield a high-purity sample of high-energy
photons that is almost immediately available (median delay of
5~hours) for real-time cross-correlation with \ice\ neutrinos.

The high suitability of the \fermi\ LAT and \ice\ neutrino datasets
for joint analysis prompted our previous archival search for
subthreshold neutrino + gamma-ray (\nugamma) emitting sources in the
\ice\ 40-string (hereafter IC40) public neutrino dataset
\citep{keivani2015}. This work, carried out under the auspices of the
Astrophysical Multimessenger Observatory Network (AMON\footnote{AMON
  website: \url{http://www.amon.psu.edu/}};
\citealt{amondesg,amonrt}), calculated pseudo-likelihoods for all
candidate \nugamma\ pairs and compared the cumulative distribution of
this test statistic to a null distribution derived from scrambled
datasets (using the Anderson-Darling test; \citealt{adtestpaper}). The
sensitivity of the analysis was calibrated via signal injection,
allowing a rough mapping of Anderson-Darling $p$-value to the number
of injected pairs. While the observed test statistic distribution, and
the $p$-value of 4\% versus the null distribution, were consistent
with the presence of $\approx$70 signal pairs out of 2138 observed
coincidences, subsequent vetting tests provided no reason to suspect
the presence of a cosmic signal.

The present work can be considered, in part, a continuation and
extension of this earlier investigation. First, we revisit the IC40
analysis using the new \fermi\ Pass~8 reconstruction and extend
the analysis to the \ice\ public 59-string dataset (hereafter IC59),
which covers both Northern and Southern hemispheres. Second, we extend
the previous test statistic to incorporate the possibility of single
neutrino + multiple gamma-ray coincidences, which provides an
unbounded statistic suitable for identification of individual
high-significance events. Finally, we have divided the
\fermi\ bandpass into three energy ranges in our background
calculations, which we expect will improve the sensitivity of the
analysis for relatively hard-spectrum transients.

The paper is organized as follows: Details of the datasets are
provided in Sec.~\ref{sec:data}, while our statistical approach and
signal injection studies are discussed in
Sec.~\ref{sec:meth}. Unscrambling of the neutrino datasets and results
are presented in Sec.~\ref{sec:results}, while Sec.~\ref{sec:conc}
provides our conclusions, including suggestions for future work.


\section{Datasets}
\label{sec:data}

This analysis was performed using available \ice\ and \fermi\ LAT
public data over the period of temporal overlap between the two
observatories. The relevant \fermi\ data were the Pass~8 photon
reconstructions available from the LAT FTP server\footnote{LAT data
  located at
  \url{ftp://legacy.gsfc.nasa.gov/fermi/data/lat/weekly/photon/}}.
These photon events were filtered using the Fermi Science Tools,
keeping only photons with a zenith angle smaller than 90\arcdeg,
energies between 100~MeV and 300~GeV, detected during good time
intervals (GTI) provided in the LAT satellite
files\footnote{\fermi\ satellite files located at
  \url{ftp://legacy.gsfc.nasa.gov/fermi/data/lat/weekly/spacecraft/}}.

The point spread function (PSF) of the LAT is given by a double King
function with the parameters depending on the photon energy,
conversion type, and incident angle with respect to the LAT boresight
\citep{fermipsf}. At lower energies (hundreds of MeV), the angular
uncertainty can be several degrees, especially for off-axis photons. At
$\egam > 1$\,GeV the average uncertainty drops below 1\arcdeg, and
at $\egam \simgt 100$\,GeV angular uncertainties are better than
0.1\arcdeg.

\begin{figure}
\includegraphics[width=\columnwidth]{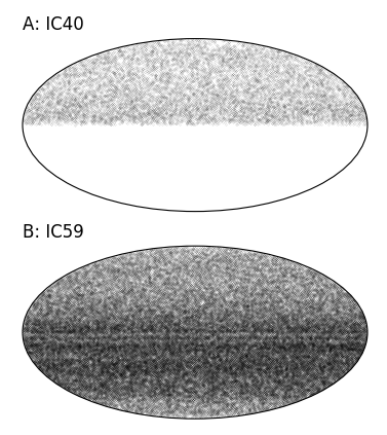}
\caption{Neutrino sky positions from IC40 and IC59. No cosmic
  structure nor significant point-source detections have been reported
  from these data \citep{ic40,ic59}.}
\label{fig:numap}
\end{figure}

Public data from the 40-string (IC40\footnote{IC40 data available at
  \url{http://icecube.wisc.edu/science/data/ic40}}) and 59-string
(IC59\footnote{IC59 data avalable at
  \url{http://icecube.wisc.edu/science/data/IC59-point-source}})
configurations of \ice\ were used \citep{ic40,ic59}. IC40 ran from
April~2008 to May~2009 and contains 12,876 neutrinos over the northern
hemisphere. This corresponds to Weeks 9 to 50 of the \fermi\ mission,
which has public data available from 4~August 2008. Applying our cuts
to the \fermi\ data yield 7.2 million northern-hemisphere photons
during IC40, and reduces the IC40 neutrino dataset to 8871 events over
the approximately nine-month period of joint operations. IC59 ran from
May~2009 to May~2010 and contains 107,569 neutrino events; this period
corresponds to weeks 50 to 104 of the \fermi\ mission, and yields
19.4~million photon events passing our cuts. Fig.~\ref{fig:numap}
shows neutrino sky maps for IC40 and IC59 in equatorial coordinates.

We adopt a Gaussian form for the \ice\ PSF. For IC40, the angular
uncertainty for each neutrino is set at
0.7\arcdeg\ \citep{icmoonshadow2013}.  Angular uncertainties are
provided for the IC59 events, and we use the reported angular
uncertainty for each event.

\begin{figure}
\includegraphics[width=\columnwidth]{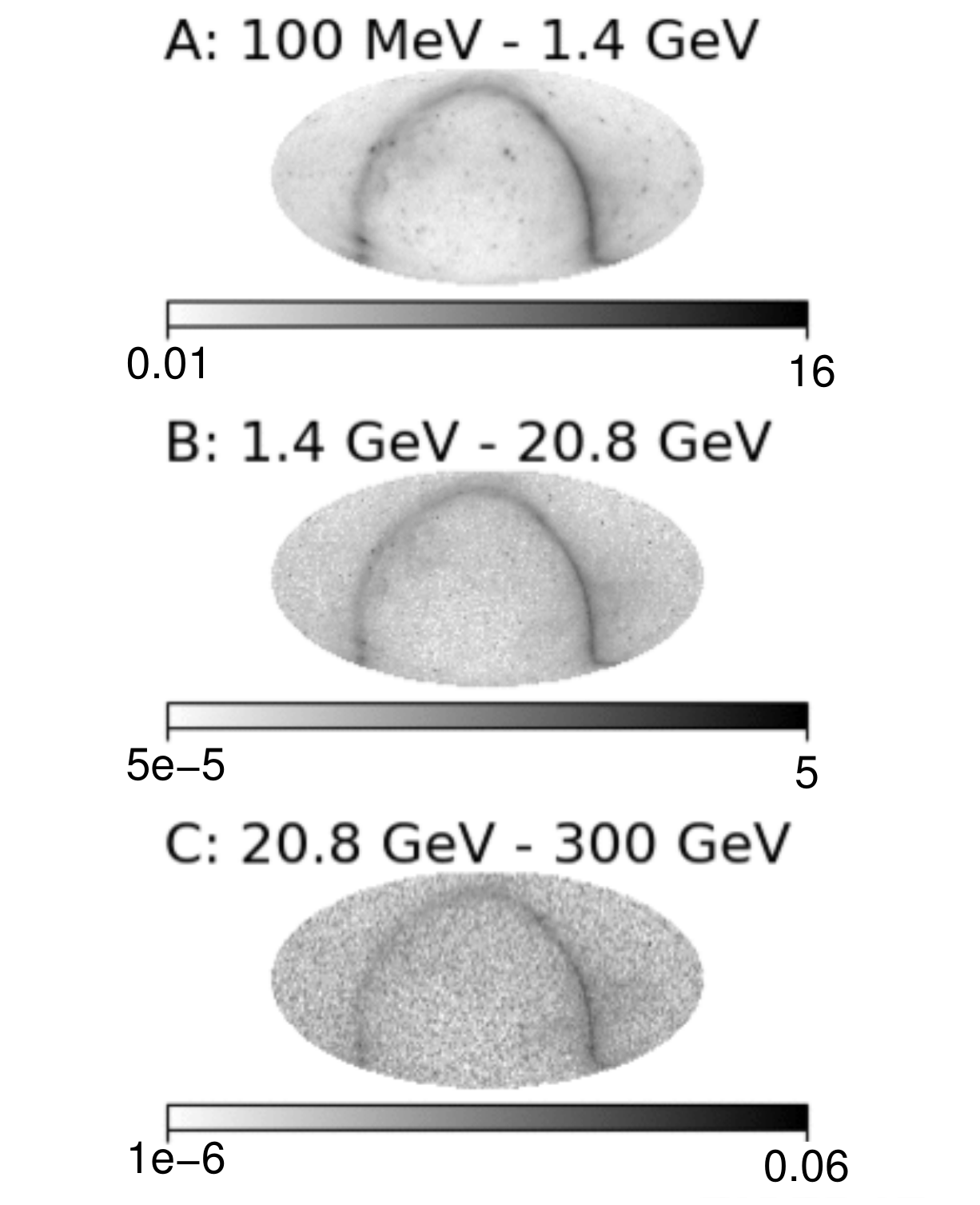}
\caption{\fermi\ LAT all-sky exposure-corrected images. We divide the
  \fermi\ data into three bins of equal width in $\log \egam$ and
  calculate mission-averaged all-sky images to determine the expected
  background rate for each photon in the primary analysis. Greyscale
  intensity units, indicated by the color bars, are photons per 200
  seconds per square meter per square degree.}
\label{fig:phbkg}
\end{figure}

A {\tt healpix} \citep{healpix} map of resolution 8 (NSide=256, mean
spacing of 0.23\arcdeg ) was constructed using the entire \fermi\ data
set (weeks 9 to 495 at the time of creation) with the same photon
selection criteria used for IC40 and IC59. Using the {\tt HEASoft}
\citep{heasoft} software, events were binned into three
logarithmically uniform energy bins. Each energy bin was then further
binned into a {\tt healpix} map, with the live time calculated via a
Monte Carlo simulation. Dividing the counts map by the live time map
produced the \fermi\ exposure map. Zero-valued (low-exposure) pixels
were replaced by the average of the nearest neighbor pixels. Our three
resulting all-sky \fermi\ maps are shown in Fig.~\ref{fig:phbkg}. Due
to the additional reconstruction uncertainty in the \fermi\ PSF for
high-inclination events (inclination angle greater than 60\arcdeg),
three additional maps were generated by further averaging all pixels
with their nearest neighbors.

\section{Methods}
\label{sec:meth}

\subsection{Significance Calculation}

Our analysis begins by filtering for coincidences between an
individual neutrino event and all photons within 5\arcdeg\ angular
separation and $\pm$100~s arrival time, as per
\citet{keivani2015}. The angular acceptance cut corresponds
approximately to the maximum 1$\sigma$ radial uncertainty for
\fermi\ LAT photons satisfying our event selection. The temporal
acceptance window is chosen to include $\approx$90\% of classical
gamma-ray bursts \citep{batse2b}. For each coincidence, a
pseudo-log-likelihood test statistic, $\lambda$, is calculated as
follows:
\begin{equation}
  \lambda= 2 \ln \frac{(P_{\gamma 1}(\vec{x}) P_{\gamma 2}(\vec{x})
    ... P_{\gamma n}(\vec{x})) n! (P_{\nu}(\vec{x}))}{B_1(\vec{x},E_1,
    \theta_1)B_2(\vec{x},E_2,\theta_2)...B_n(\vec{x},E_n,\theta_n) }
\label{eq:lambda}
\end{equation}
where $n$ is the number of photons coincident with the neutrino,
$P_{\gamma i} (\vec{x})$ is the energy-dependent point spread function
(PSF) of the LAT at the best fit position, $\vec{x}$, and
$P_{\nu}(\vec{x}))$ is the \ice\ PSF at the best fit position. Both
PSFs have units of probability per square degree. The
$B_i(\vec{x},E_i,\theta_i)$ are the LAT background terms in units of
photons per square meter (approximating the \fermi\ effective area)
per 200~s (our temporal window) per square degree for each $\gamma_i$,
given its energy $E_i$ and inclination angle $\theta_i$.  In this
metric, larger values of $\lambda$ indicate a higher-likelihood
correlated multiplet. This pseudo-log-likelihood statistic is the
natural extension of the \citet{keivani2015} test statistic to
multi-photon coincidences, via the Poisson likelihood of generating an
$n$-fold coincidence from background; in the prior approach, each
\nugamma\ coincidence was treated separately.

The best fit position $\vec{x}$ is determined as the location of
maximal PSF overlap. As the overlap of a double King function with a
Gaussian function cannot be solved analytically, the best-fit position
is found numerically. For single neutrino + mutiple photon
coincidences, the event photon multiplicity is determined by
optimization: We compare the $\lambda$ value at maximum multiplicity
to that which would result if the photon with the lowest PSF density
at the best-fit position were excluded (after recalculating the
best-fit position and $\lambda$), and iteratively exclude photons
until $\lambda$ no longer increases.


\subsection{Analysis Definition}
\label{sub:search}

We generate a set of 10,000 Monte Carlo scrambled versions of each of
our three datasets in order to characterize their null distributions
and define analysis thresholds, prior to performing any analysis of
the unscrambled datasets. Our scrambling procedure begins by shuffling
the full set of neutrino detections, associating each original
neutrino $\nu_i$ with another randomly selected neutrino $\nu_j$. Each
neutrino $\nu_i$ retains its original declination and angular error
and receives the original arrival time of neutrino $\nu_j$, with its
new right ascension derived by adjusting the original right ascension
for the difference in local sidereal time between the original and new
arrival times, the same approach as in
\citet{turleyblazar}. \fermi\ LAT photons are not scrambled as the LAT
data contains known sources and extensive (complex)
structure. Coincidence analysis is carried out for each scrambled
dataset and $\lambda$ values are calculated for the resulting
\nugamma\ coincidences via Eq.~\ref{eq:lambda}.

This analysis presents two discovery scenarios. First, since our test
statistic $\lambda$ is unbounded, the null distribution provides us
with threshold values which can be used to identify
individually-significant coincidences and estimate their false alarm
rates. We define two such thresholds, \lamten, the value exceeded (one
or more times) in 1 of 10 scrambled datasets, and \lamhund, the value
exceeded in 1 of 100 scrambled datasets. For analyses treating a year
of observations, these two thresholds would correspond to events with
false alarm rates of 1~decade$^{-1}$ and 1~century$^{-1}$,
respectively.

Under this approach (and without accounting for the trials factor, see
below), observation of a single event above \lamhund, or two events
above \lamten, would constitute evidence of joint \nugamma\ emitting
sources, while observation of two events above \lamhund\ or four
events above \lamten\ would enable a discovery claim.

In the absence of any individually-significant events, there remains
the opportunity for discovery of a subthreshold population of
\nugamma\ emitting sources. By design, true cosmic coincidences are
biased to higher $\lambda$ values (Fig.~\ref{fig:hists}), and a
population containing a sufficient number of such signal events can be
distinguished from the null distribution using the Anderson-Darling
$k$-sample test.  The $k$-sample test is used to establish mutual
consistency among $k$ observed datasets ($k=2$ for a two-sample test),
testing against the null hypothesis that they are drawn from a single
underlying distribution.

Given our choice to make two statistical tests on each of three
predefined datasets (IC40, \icfnorth, and \icfsouth), we apply an
$N_{\rm trials}=6$ trials penalty to our unscrambled analyses
(Sec.~\ref{sub:unscrambled}).


\subsection{Signal Injection}
\label{sub:injection}

To estimate our sensitivity to cosmic \nugamma\ emitting source
populations, we generate signal-like events and inject these into
scrambled datasets, comparing the results to the null distribution. 

Since we test for $\gamma$ multiplicity, as part of this process we
must adopt a procedure for determining whether each injection is a
$\gamma$ singlet, doublet, or $n_\gamma > 2$ multiplet. To determine
the appropriate $n_\gamma$ distribution, we assume a population of
sources emitting one neutrino, with associated photon fluence
distributed according to $N(S\ge S_0) \propto S_{0}^{-3/2}$, where
$S_0$ is a threshold photon fluence and $N(S\ge S_0)$ is the number
of events observed with fluences greater than or equal to this
threshold; we note that an $S_{0}^{-3/2}$ dependance is expected for
source populations of arbitrary luminosity function distributed in
Euclidean space.

Adopting a minimum considered fluence of $S_{\rm min}=0.001$~photons,
and inverting this relation, we generate the expectation value for the
multiplicity of any event as $\ngexp = S_{\rm min}\, u^{-2/3}$, where
$u$ is a uniform random variable. We then generate the observed
$n_\gamma$ by drawing randomly from the Poisson distribution with
expectation value \ngexp.  Excluding zero-multiplicity events, we are
left with the following $n_\gamma$ distribution: 93\% singlet, 4.8\%
doublet, 1.1\% triplet, 0.5\% quadruplet, 0.3\% quintuplet, 0.2\%
sextuplet, and 0.1\% septuplet (the highest multiplicity we allow). As
an aside, we note that this is (approximately) the unique and
universal distribution expected to arise in these cases, for
extragalactic source populations extending to modest redshift
($z\simlt 1$) with weak evolution, and thus distributed in
near-Euclidean space. 

A signal event of photon multiplicity $n_\gamma$ is generated by
centering the PSF for $n_\gamma$ LAT photons and an \ice\ neutrino
at the origin. The neutrino localization uncertainty is drawn from the
full set of \ice\ neutrino uncertainties, while the inclination
angles and conversion types of the photons are drawn from the full set
of these distributions within the \fermi\ dataset. Photon energies are
drawn from a power law with a photon index $\Gamma=-2$. The photons
and the neutrino are placed randomly according to their respective
PSFs. A random sky position is then chosen as the best fit position
for this coincidence, and a $\lambda$ value is calculated following
the methods of Sec.~\ref{sec:meth}. Since the $\lambda$ calculation
involves maximizing $\lambda$ by exclusion of outlying photons, many
events end up with some of the injected photons excluded. Cumulative
distributions for the null and signal-only distributions are shown in
Fig.~\ref{fig:hists}.

\begin{figure}
\includegraphics[width=\columnwidth]{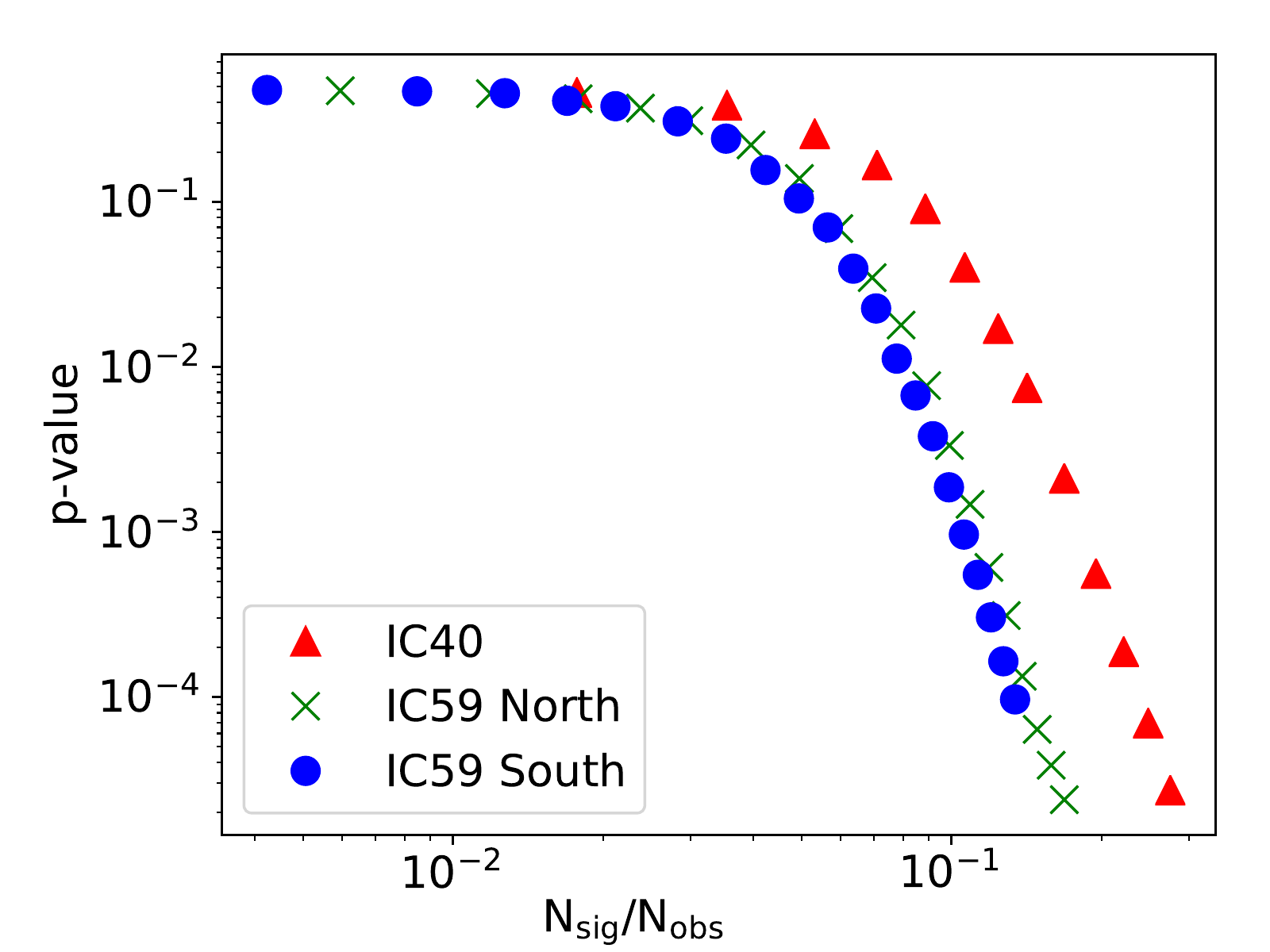}
\caption{Analysis sensitivity, plotted as Anderson-Darling two-sample
  $p$-value versus fraction of coincidences that result from signal
  events, \ninj/\nobs. Results for IC40 are plotted in red,
  \icfnorth\ in green, and \icfsouth\ in blue. As expected, these
  sensitivities scale roughly as $1/\sqrt{\nobs}$, so that the larger
  \icfnorth\ and \icfsouth\ datasets provide superior fractional
  sensitivity than IC40. }
\label{fig:ad}
\end{figure}

To calculate the sensitivity of our analysis, we inject an increasing
number of signal events \ninj\ into a scrambled (null) distribution
and compare the signal-injected and null $\lambda$ distributions using
the \adstat\ $k$-sample test. We carry out 10,000 trials for each
selected value of \ninj\ and plot the mean resulting $p$-value against
\ninj, for each of our datasets, in Fig.~\ref{fig:ad}. In this way we
estimate the threshold value of \ninj\ that is required to yield a
statistically-significant deviation from the null distribution for
each of the datasets ($n_{\rm inj,1\%}$ and $n_{\rm inj,0.1\%}$
columns in Table~\ref{tab:res}).


\subsection{Analysis Sensitivity and Expectations}
\label{sub:scrambled}


\begin{figure}
\includegraphics[width=\columnwidth]{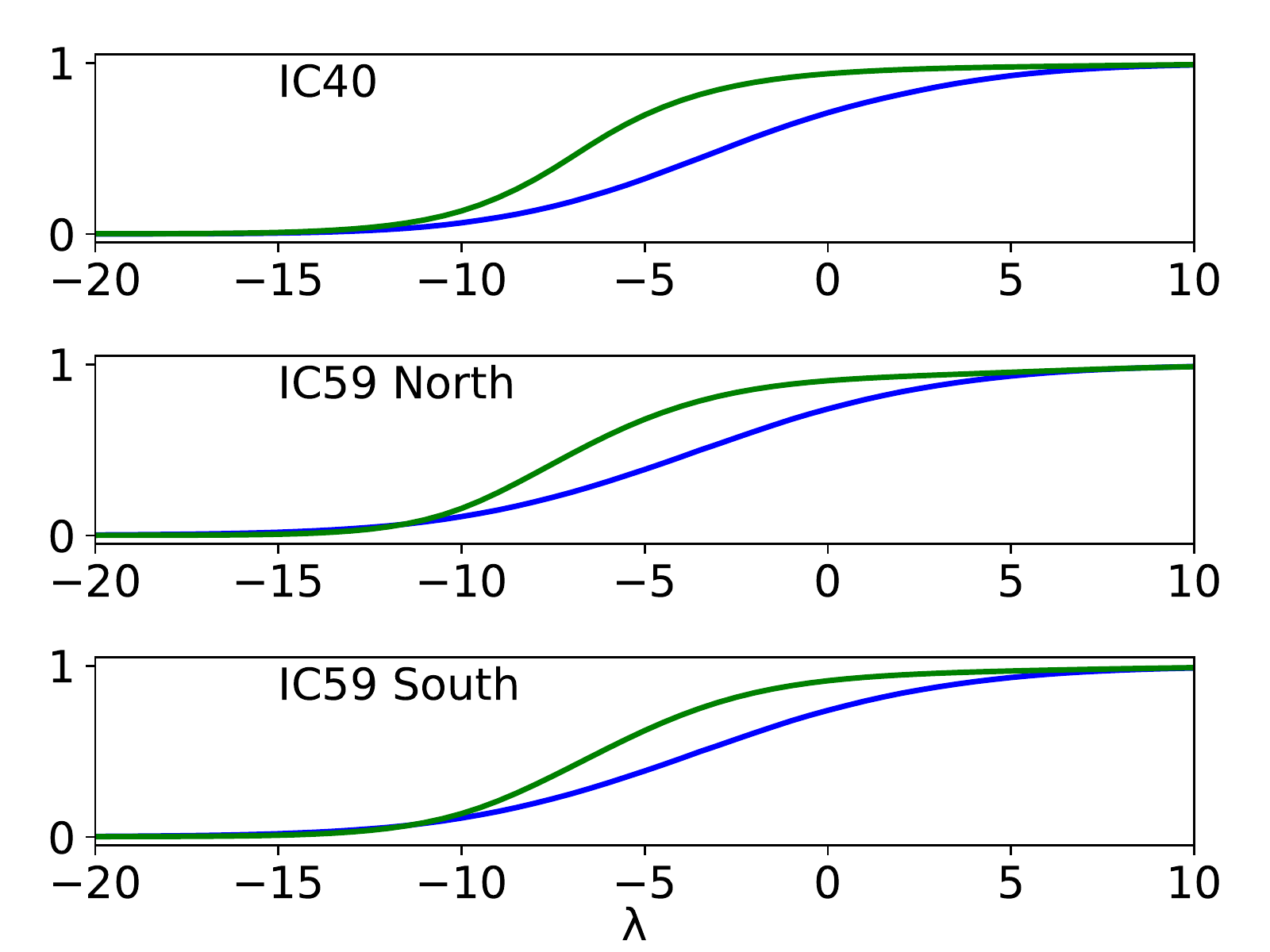}
\caption{Cumulative distribution of pseudo-log-likelihood ($\lambda$)
  values from null/scrambled (green) and signal-only (blue)
  realizations of the IC40 (top), \icfnorth\ (middle), and
  \icfsouth\ (bottom) datasets. Note that the tails of the
  distributions extend far off of the plots}
\label{fig:hists}
\end{figure}

Carrying out the scrambled analysis on the IC40 and IC59 data produces
the null $\lambda$ distributions shown in Fig.~\ref{fig:hists}. Key
statistics from these analyses are summarized in
Table~\ref{tab:res}. Most of the simulated events with $\lambda >
\lamhund$ in \icfnorth\ scrambled runs result from a scrambled
neutrino landing in near coincidence with one of two GRBs detected by
the LAT during our period of observation. GRB\,090902B
\citep{latgrb090902} placed over 200 photons on the LAT, giving a
maximum $\lambda = 2560.2$ in a 218-photon coincidence. GRB\,100414A
\citep{latgrb100414} placed over 20 photons on the LAT and yields a
maximum $\lambda = 91.2$ in a 10-photon coincidence. Excluding all
coincidences with either of these GRBs would yield a threshold of
$\lamhund = 35$ for the \icfnorth\ data, rather than the GRB-inclusive
value of $\lamhund = 49.0$.

Given the number of signal-like \nugamma\ required to yield a $p<1\%$
deviation in the \adstat\ $k$-sample test, we estimate our analysis
would detect $>$150 source-like \nugamma\ coincidences from IC40
($>$13\% of the total number of coincidences in IC40), $>$440 from
\icfnorth\ ($>$9\%), and $>$565 from \icfsouth ($>$9\%); see
Table~\ref{tab:res}.

\begin{deluxetable*}{lrrrrrrrr}

  \tablecolumns{9}
  \tablecaption{Coincidence search results\label{tab:res}}

  \tablehead{%
    \colhead{~} &
    \colhead{~} &
    \multicolumn{4}{c}{Thresholds} &
    \multicolumn{3}{c}{Observed} \\
    \colhead{Dataset} &
    \colhead{$\langle n_{\nu+\gamma} \rangle$} &
    \colhead{\lamten} &
    \colhead{\lamhund} &
    \colhead{$n_{\rm inj,1\%}$} &
    \colhead{$n_{\rm inj,0.1\%}$} &
    \colhead{$n_{\nu+\gamma}$} &
    \colhead{$\lambda_{\rm max}$} &
    \colhead{$p_{\rm A-D}$}} 
  
  \startdata
  IC40       & $1090 \pm 30$ & 23.9 & 27.2 & 150 & 210 & 1128 & 20.3 &  63\%  \\
  IC59-North & $4970 \pm 65$ & 26.5 & 49.0 & 440 & 570 & 5046 & 17.8 & 16.8\% \\
  IC59-South & $7072 \pm 76$ & 26.8 & 31.5 & 565 & 740 & 7080 & 24.4 &  3.8\% \\
  \enddata

  \tablecomments{$\langle n_{\nu+\gamma} \rangle$ is the expected number of
    neutrinos observed in coincidence with one or more gamma-rays, as
    derived from 10,000 Monte Carlo scrambled realizations of each
    dataset. \lamten\ and \lamhund\ are the thresholds above which a
    coincidence is only observed once per 10 or 100 scrambled
    datasets, respectively. $n_{\rm inj,1\%}$ and $n_{\rm
      inj,0.1\%}$ are the number of injected signal events required
    in simulations to give an Anderson-Darling test statistic of
    $p<1\%$ and $p<0.1\%$, respectively, by comparison to the null
    distributions for each dataset. $n_{\nu+\gamma}$ is the number of neutrinos
    observed in coincidence with one or more gamma-rays in the
    unscrambled data, $\lambda_{\rm max}$ is the maximum observed
    $\lambda$ for each dataset, and $p_{\rm A-D}$ is the value of the
    Anderson-Darling test statistic from comparison of the observed
    $\lambda$ distribution to its associated null distribution.}

\end{deluxetable*}


\section{Results}
\label{sec:results}

\subsection{Coincidence Search}
\label{sub:unscrambled}

\begin{figure}
\includegraphics[width=\columnwidth]{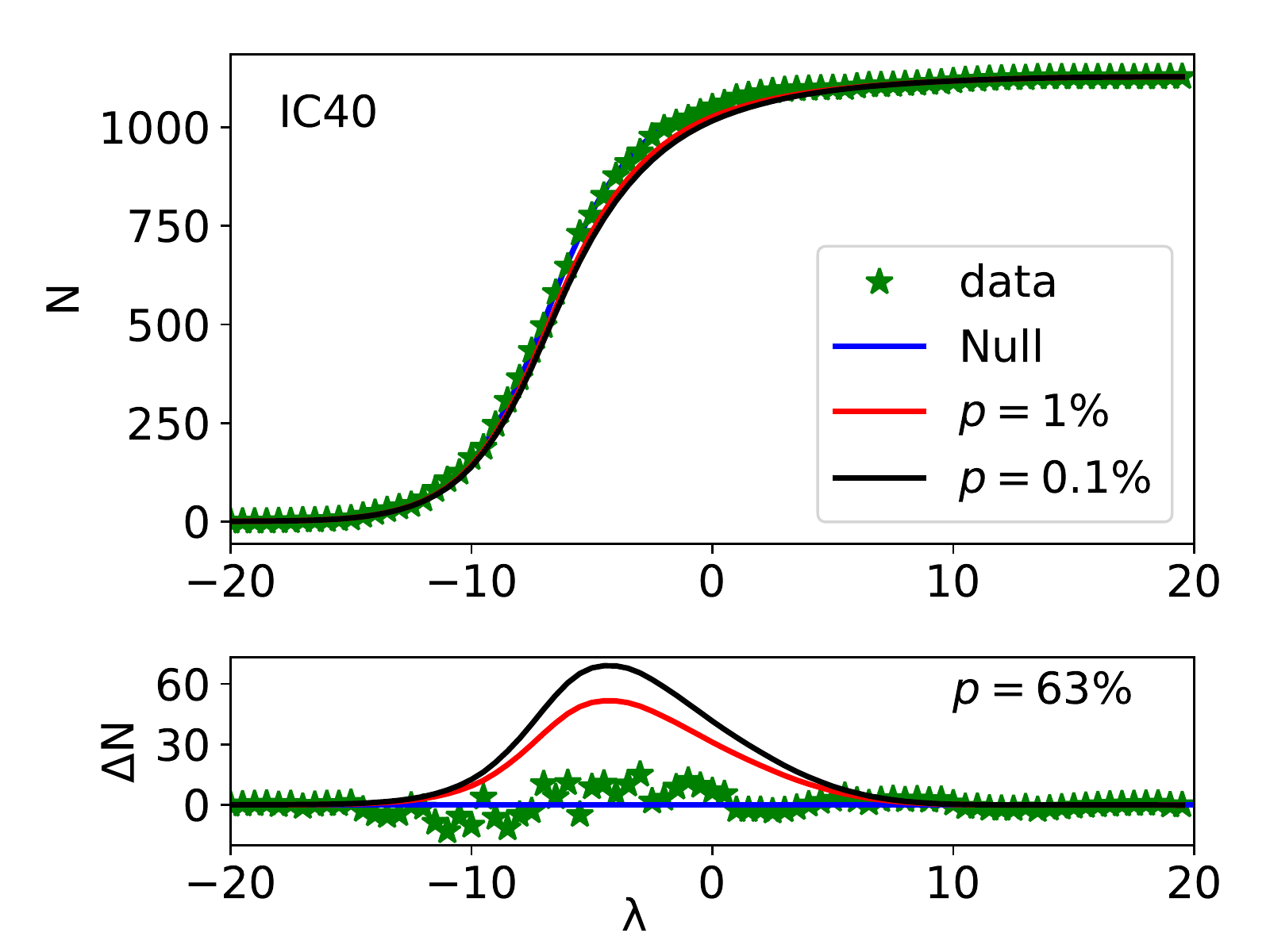}
\caption{Cumulative and residual test statistic ($\lambda$)
  distributions for IC40 ($n_{\nu+\gamma}=1128$). Upper panel:
  Cumulative IC40 $\lambda$ distributions for unscrambled data (green
  stars), scrambled data / null distribution (blue line), and
  signal injections yielding $p=1\%$ (red line) and $p=0.1\%$ (black
  line). Lower panel: Residuals, plotted as null minus alternative,
  for IC40 data (green stars) and the two signal injection
  distributions (red and black lines).\label{fig:res40}}
\end{figure}

\begin{figure*}
\includegraphics[width=\columnwidth]{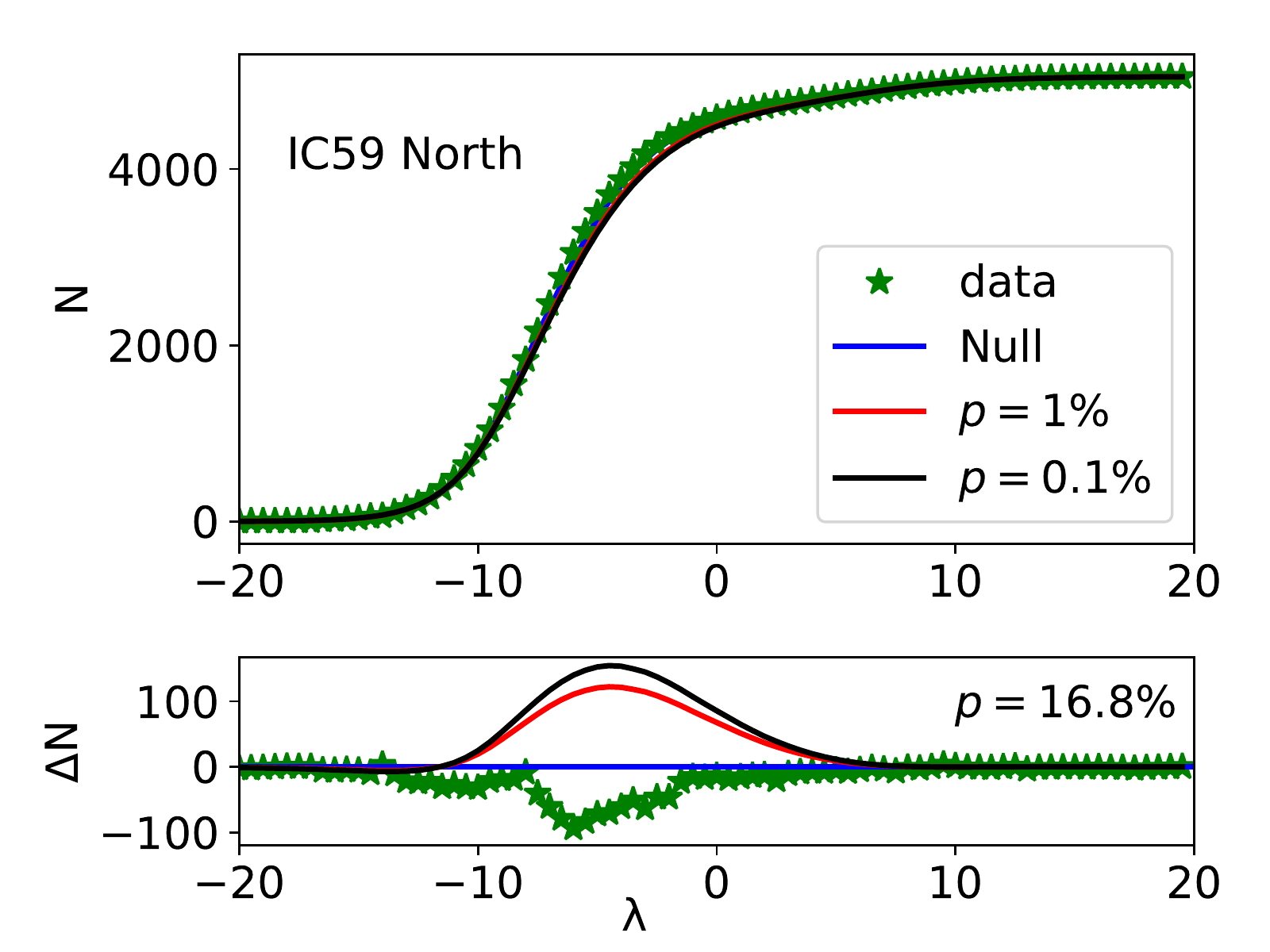}
\includegraphics[width=\columnwidth]{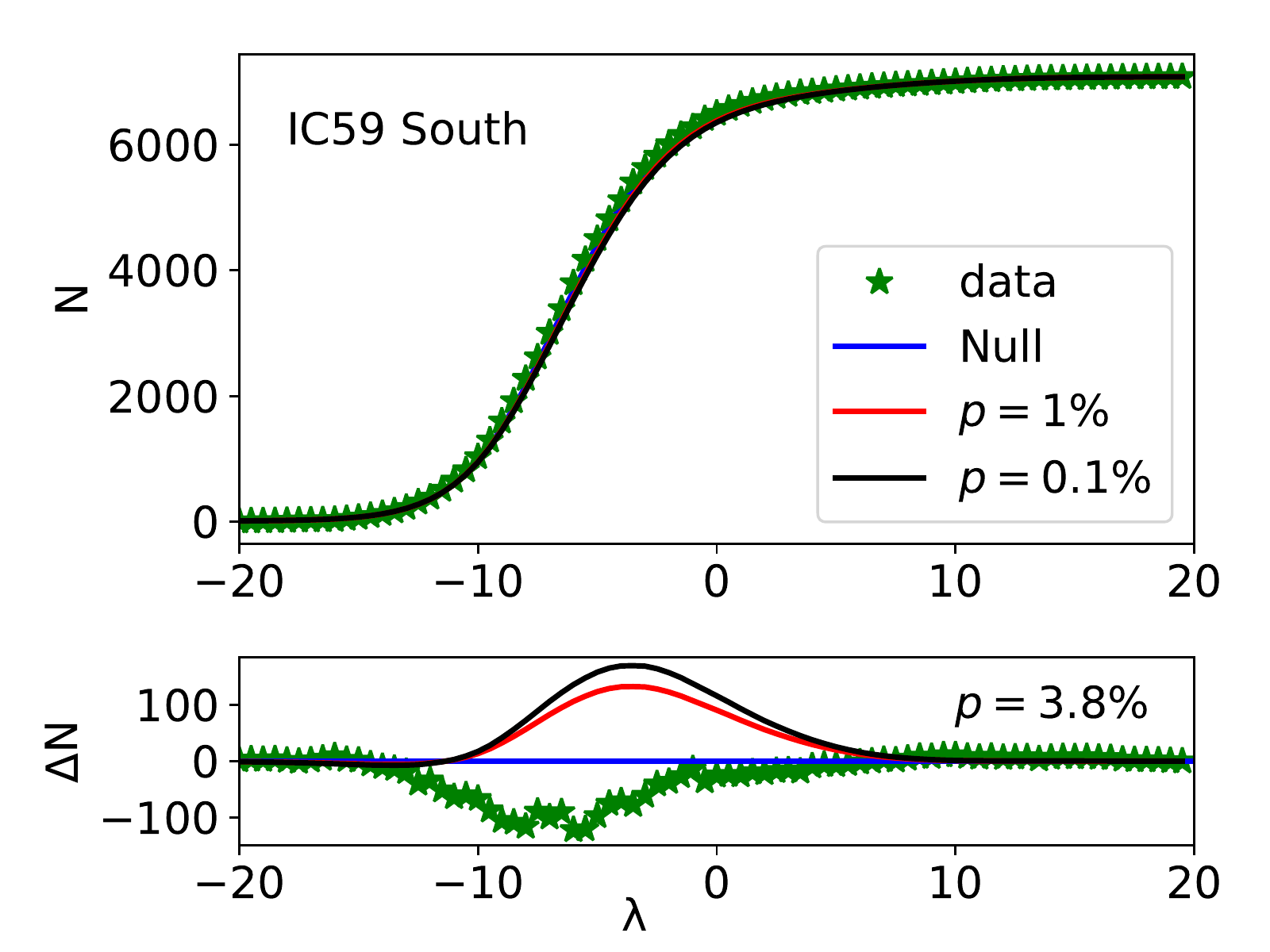}
\caption{Cumulative and residual test statistic ($\lambda$)
  distributions for \icfnorth\ (left) and \icfsouth\ (right). 
  Left: Cumulative (upper panel) and residual (lower panel) $\lambda$
  distributions for \icfnorth\ ($n_{\nu+\gamma}=5046$), including
  unscrambled data (green stars), scrambled data / null distribution
  (blue line), and signal injections yielding $p=1\%$ (red line) and
  $p=0.1\%$ (black line). Residuals are plotted as 
  null minus alternative. 
  Right: Cumulative (upper panel) and residual (lower panel) $\lambda$
  distributions for \icfsouth\ ($n_{\nu+\gamma}=7080$), including
  unscrambled data (green stars), scrambled data / null distribution
  (blue line), and signal injections yielding $p=1\%$ (red line) and
  $p=0.1\%$ (black line).
  A similar and unexpected pattern is noted
  in the residual $\lambda$ distributions for the \icfnorth\ and 
  \icfsouth\ datasets. \label{fig:res59}} 
\end{figure*}

Applying our analysis to the three unscrambled neutrino datasets
yields the results summarized in Table~\ref{tab:res}.
Fig.~\ref{fig:res40} and Fig.~\ref{fig:res59} show the $\lambda$
distributions for the unscrambled data for IC40, \icfnorth, and
\icfsouth, along with the null distributions, and distributions for
signal injections yielding $p$-values from the Anderson-Darling test
of 1\% and 0.1\%, respectively. All distributions are normalized to
the number of coincidences $n_{\nu+\gamma}$ observed in the
unscrambled data.  No $\lambda$ values were detected above the
\lamten\ threshold in any of the analyses.  Notably, as seen in
Fig.~\ref{fig:res59}, the \icfnorth\ and \icfsouth\ data show an
excess of lower $\lambda$ values by comparison to the null
distributions, unlike the excess of higher $\lambda$ values expected
from a signal population.

Given our six trials, a minimum observed single-trial $p=3.8\%$
corresponds to a trials-corrected value of $p_{\rm post}=20.7\%$,
which is not significant. However, the similar scale and shape of the
residual patterns for \icfnorth\ and \icfsouth\ lead us to seek out
possible causes of these residual patterns. To illustrate this point,
combining $p$-values from these two datasets (our most sensitive)
by Fisher's method gives a joint $p=3.9\%$ (single trial) as the
probability of generating two such large deviations by random chance.

We note that we have not conceived of any way for systematic effects
to generate the IC59 residuals and low $p$-values, simply because any
errors or simplifications in the analysis (which certainly exist) are
replicated across all scrambled datasets. Rather, the only way to
generate these effects (if they are not due to random statistical
fluctuation) is via spatio-temporal correlation of neutrinos and
gamma-rays. Such correlation would imply either cosmic sources, or at
a minimum, correlated emission (e.g., enhancement toward the Galactic
plane or Supergalactic plane) and hence, require structure in the
neutrino sky which has not previously been observed.

We divide our further explorations into two approaches: First
(Sec.~\ref{sub:vetting}), we further vet the IC59 data against our
original hypothesis, to test for any evidence that short-duration
($\delta t < 100$\,s) \nugamma\ transients are really present in the
data. Second (Sec.~\ref{sub:correlations}), we test for
longer-duration \nugamma\ spatio-temporal correlations that might
have an effect on our original analysis. 


\subsection{IC59 Vetting}
\label{sub:vetting}

We vet the IC59 datasets to evaluate whether the low $p$-values and
systematic trends in the residual $\lambda$ distributions that we
observe could be due to \nugamma\ transient sources, as per our
original hypothesis, but below the sensitivity of those analyses. We
exclude the IC40 dataset from these tests as it is less sensitive to
the presence of cosmic sources (see Fig.~\ref{fig:ad} and
Sec.~\ref{sub:scrambled}).

Specifically, we check for systematic trends or anomalies in the
spatial and temporal distributions of the neutrino-coincident photons
that might account for the unexpected deviation to lower $\lambda$
values. We test separately for deviations in the distributions of the
photons' angular and temporal separations from their coincident
neutrino, for \icfnorth\ and \icfsouth.


With regards to the angular separation distributions, we note that a
systematic underestimation of \ice\ neutrino localization
uncertainties might cause suppressed $\lambda$ values relative to
simulations, due to the $P_\nu$ term in the pseudo-likelihood. In a
similar vein, even if all neutrino and gamma-ray localization
uncertainties are accurately characterized, a systematically softer
spectrum for cosmic \nugamma\ sources ($\Gamma < -2$) would suppress
$\lambda$ values via the $P_{\gamma i}$ terms in the
pseudo-likelihood, since higher-energy LAT photons are better
localized.

We construct five-bin histograms of the angular separations of all
neutrino-associated photons, with bin boundaries chosen to make the
null distribution approximately flat (equal numbers of photons in each
bin). We then calculate the $\chi^2$ statistic for the unscrambled
data compared to the (flat) null distribution.


\begin{figure}
\includegraphics[width=\columnwidth]{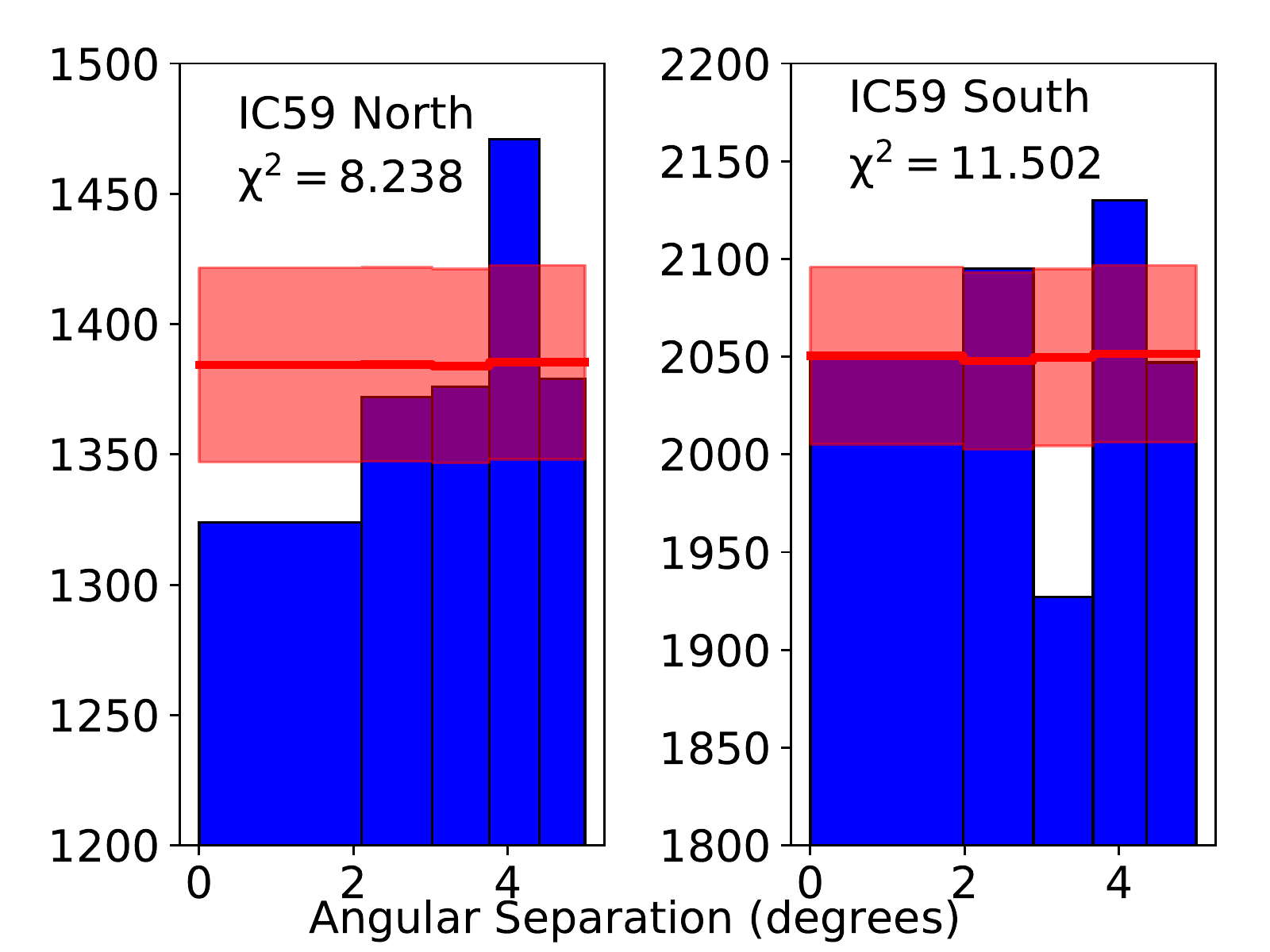}
\caption{IC59 $\nugamma$ angular separations. We test the observed
  angular separation distribution for neutrino-coincident photons
  (blue histograms) in \icfnorth\ (left) and \icfsouth\ (right)
  against the null distribution (red), which is approximately flat by
  construction (via choice of histogram bin boundaries). The
  $\pm$1$\sigma$ ranges expected for a single dataset on the basis of
  Poisson uncertainties are indicated as the red range; note
zero-suppressed $y$ axis.\label{fig:ssep}}
\end{figure}

Fig.~\ref{fig:ssep} (note zero-suppressed $y$ axis) presents our
results. Observed \icfnorth\ angular separations are consistent with
the (flat) null distribution, exhibiting $\chi^2 = 8.238$ for 4
degrees of freedom ($p=14.3\%$).  \icfsouth\ angular separations, by
contrast, show a substantial deficit in the bin at
$\delta\theta\approx 3\arcdeg$ (and modest excesses in the bins to
either side), which results in $\chi^2=11.502$ for 4 degrees of
freedom, giving $p=4.2\%$. While this deviation is moderately
surprising, the absence of any systematic trend to low or high
separations suggests it is likely not responsible for the observed
deviation in the $\lambda$ distribution.

Here we note that a systematic trend to small angular separations
would suggest the presence of \nugamma\ sources as per our test
hypothesis, while a systematic trend to large angular separations
would suggest the presence of \nugamma\ sources with underestimated
localization uncertainties or soft gamma-ray spectra. Neither such
trend is observed.


We execute a similar analysis of the temporal separations of
coincident photons. In contrast to the angular separations, which are
incorporated into our pseudo-likelihood calculation, temporal
separations are not considered (apart from the predefined acceptance
window), so this analysis serves as an independent test of our
original hypothesis. For purposes of trials correction of any
subsequent statistics, we therefore add two trials, giving $N_{\rm
  trials}=8$.


\begin{figure}
\includegraphics[width=\columnwidth]{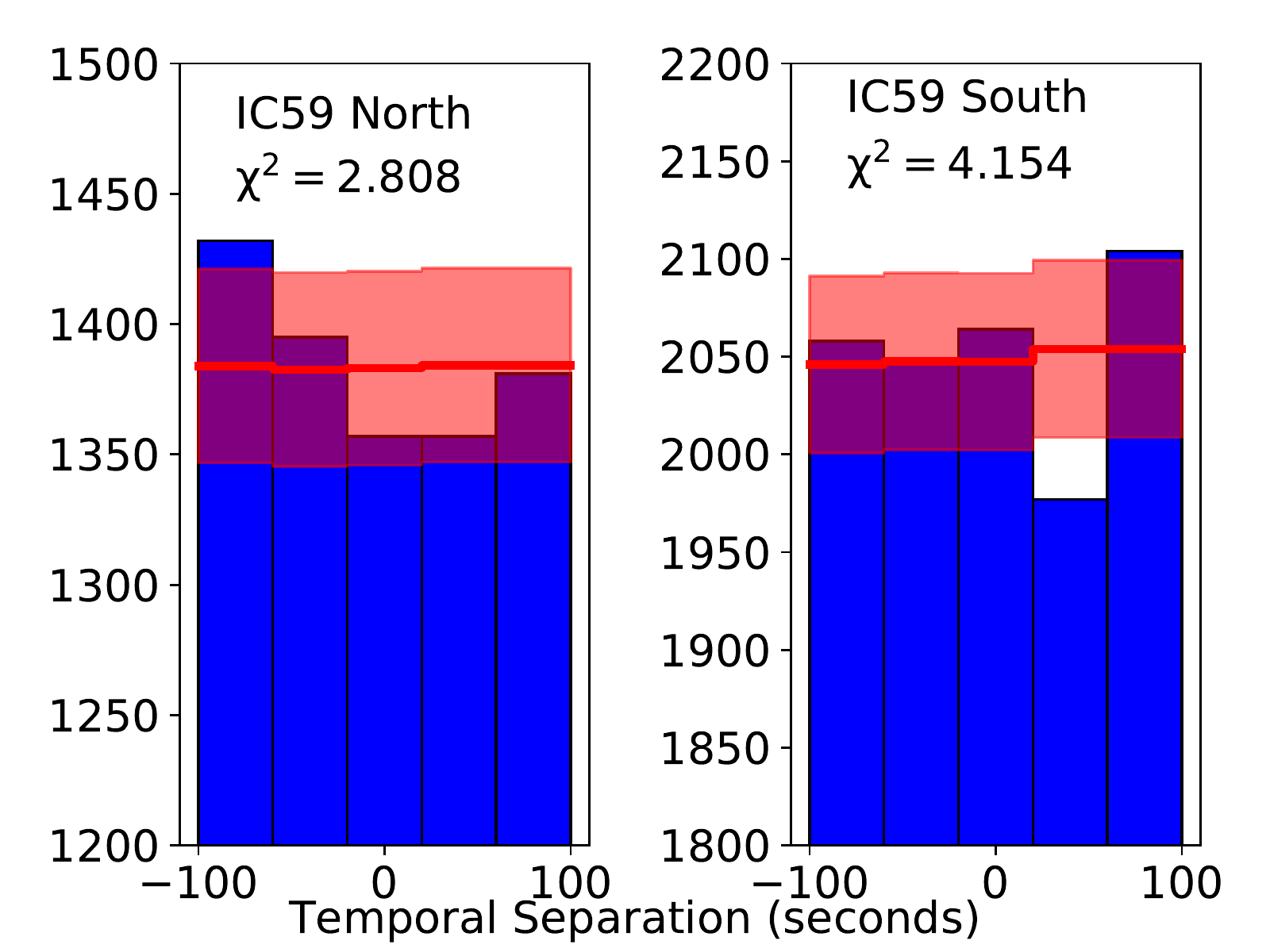}
\caption{IC59 $\nugamma$ temporal separations. We test the observed
  temporal separation distribution for neutrino-coincident photons
  (blue histograms) in \icfnorth\ (left) and \icfsouth\ (right)
  against the approximately flat null distribution
  (red). The $\pm$1$\sigma$
  ranges expected for a single dataset on the basis of 
  Poisson uncertainties are indicated by the red range; note 
  zero-suppressed $y$ axis.\label{fig:tsep}}
\end{figure}

Fig.~\ref{fig:tsep} (note zero-suppressed $y$ axis) shows our
results for temporal separations data in the two IC59
datasets. Neither dataset shows evidence for deviation from the
expected flat distribution (illustrated using scrambled datasets),
with $\chi^2$-derived $p$-values of $p=73\%$ for \icfnorth\ and
$p=53\%$ for \icfsouth.

Examining the angular and timing separations of the neutrinos and
coincident photons at higher resolution thus provides no support for
the presence of short-duration ($\delta t \simlt 100$\,s)
\nugamma\ emitting cosmic sources as conceived in our original
hypothesis. Since these are not seen, we move on to examine
alternative models that might generate \nugamma\ spatio-temporal
correlations in the data. 


\subsection{Tests for \nugamma\ Correlation}
\label{sub:correlations}

We carry out two tests for spatio-temporal correlations between the
neutrino and gamma-ray datasets beyond our original $\pm$100~s
temporal acceptance window.

First, a correlation between neutrino and photon positions on the sky,
without any temporal correlation (i.e.\ in steady state) could
suppress $\lambda$ values relative to the null hypothesis, due to the
$B_i$ gamma-ray background terms in our pseudo-likelihood
(Eq.~\ref{eq:lambda}).  

\begin{figure}
\includegraphics[width=\columnwidth]{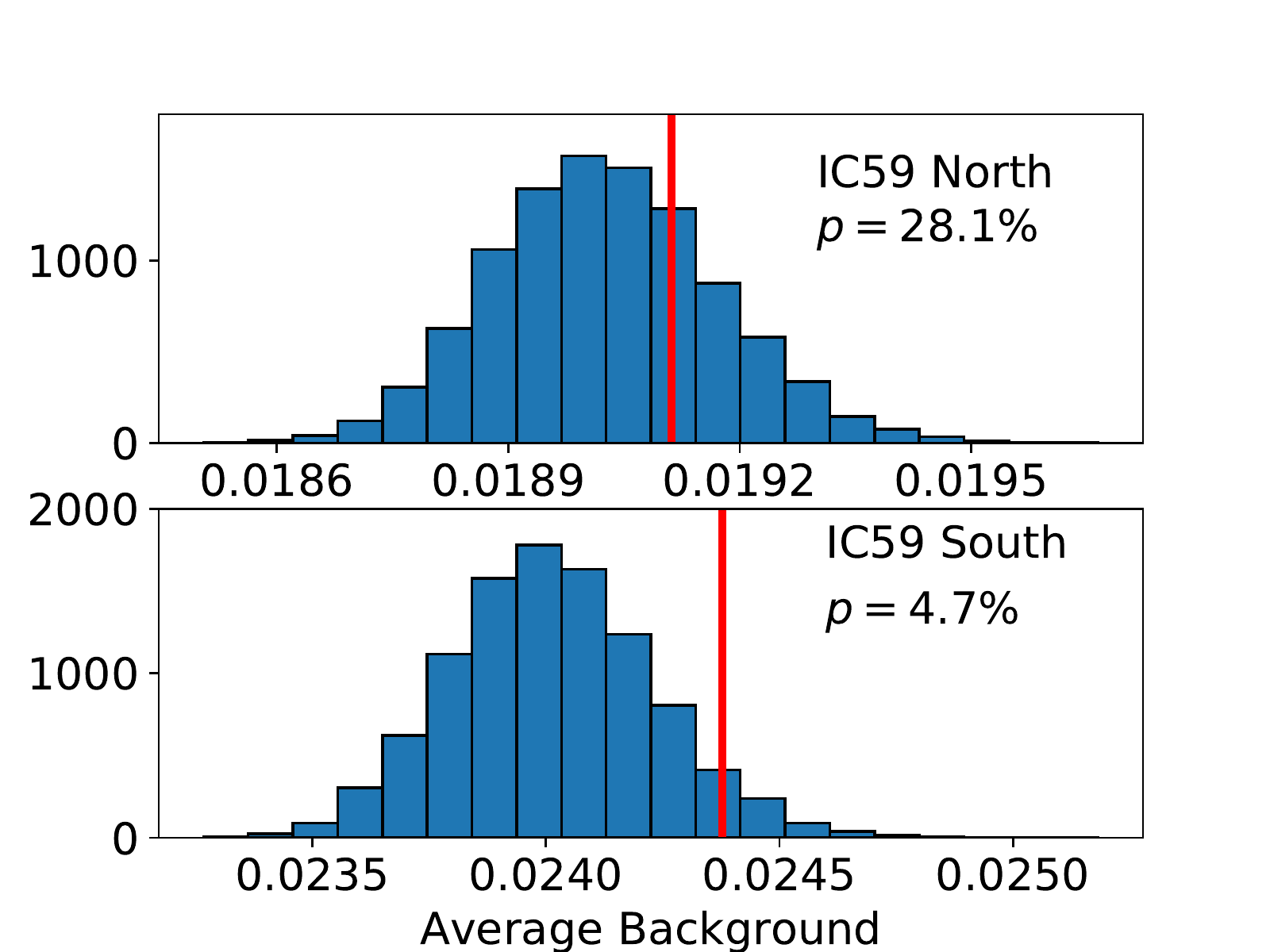}
\caption{Average \fermi\ gamma-ray background rates at the positions
  of \icfnorth\ (upper panel) and \icfsouth\ (lower panel)
  neutrinos. In each panel, the histogram shows the distribution from
  10,000 Monte Carlo scrambled datasets, while the red line marks the
  observed background rate for unscrambled data. Background rates are
  expressed in units of photons m$^{-2}$ deg$^{-2}$. Observed
  neutrino positions show a mild statistical preference for
  higher-background regions of the \fermi\ gamma-ray sky, with a joint
  $p$-value for the two hemispheres of $p=7.0\%$ by Fisher's
  method.\label{fig:bcor}}
\end{figure}

To test for positional correlation, we first construct a single
\fermi\ background map covering the full energy range. We then measure
the background value at the location of every \ice\ neutrino in
unscrambled data and compute the average photon background for the
neutrino map. This average is then compared to the average backgrounds
from each of the 10,000 scrambled datasets. The scrambled datasets
give an average background of $(1.90 \pm 0.015) \times 10^{-2}$\,
photons\, deg$^{-2}$\,m$^{-2}$ per 200\,s for \icfnorth\ and $(2.40
\pm 0.022) \times 10^{-2}$ photons\, deg$^{-2}$\,m$^{-2}$ per 200\,s
for \icfsouth. The observed average backgrounds (in the same units)
from unscrambled data are $1.91 \times 10^{-2}$ (+0.58$\sigma$;
$p=28.1\%$) for \icfnorth\ and $2.44\times 10^{-2}$ (+1.67$\sigma$;
$p=4.7\%$) for \icfsouth. These observed values are presented in the
context of the distributions from scrambled data in
Fig.~\ref{fig:bcor}.

This analysis is not an independent test for the presence of
\nugamma\ sources, but rather, an attempt to identify an underlying
reason for the trend in $\lambda$ residuals seen in
Sec.~\ref{sub:unscrambled}. Since the $p$-value for the separate
analyses, as well as their combination ($p=7.0\%$ by Fisher's method),
are within a factor of two of the $p$-values from the corresponding
$\lambda$ distribution tests, this provides reason to interpret the
latter result as (at least in part) due to the observed tendency of
IC59 neutrinos to land on systematically brighter regions of the
gamma-ray sky. We reiterate that while this tendency is present in the
data, it is not sufficiently strong (in a statistical sense) to
support an evidence claim.


As an alternative approach, we check for correlated
\nugamma\ variability on time scales beyond our predefined
$\pm$100-second temporal window but shorter than the full extent of
the \fermi\ mission. To this end, for each neutrino in our scrambled
IC59 datasets, we use the total \fermi\ mission background map to
calculate the number of photons expected to arrive within 5\arcdeg\ of
the neutrino position and $\pm$50,000~s of the neutrino arrival time
(excluding the $\pm$100~s window used in the original analysis). We
then count the number of photons arriving within this spatio-temporal
window (again, summing results across our three \fermi\ energy
bands). For each neutrino, the observed number of photons within the
extended temporal window is expressed as a Poisson fluctuation on the
number expected by normalizing against the full \fermi\ mission. We
quantify the magnitude of this fluctuation as a $p$-value and find the
equivalent number of $\sigma$ for a Gaussian distribution, yielding a
statistic that we call the local excursion $E$ for that neutrino. The
distribution of excursions from all neutrinos in unscrambled data can
then be compared to expectations from scrambled data.

We perform the same two tests that we developed in our primary
analysis above: First, we check for individual events that exhibit an
unusually large excursion, exceeding either the 1 in 10 ($E_{\times
  10}$) or 1 in 100 ($E_{\times 100}$) thresholds from scrambled
data. Second, we compare the excursion distribution from unscrambled
data to the null distribution from scrambled data using the
Anderson-Darling $k$-sample test. Since this analysis involves two
further independent tests of the two datasets, we add four trials
for purposes of trials correction of any subsequent statistics, giving
$N_{\rm trials}=12$.

Excursion thresholds for the two datasets are $E_{\times 10} = 614$
and $E_{\times 100} = 1285$ for \icfnorth, and $E_{\times 10} = 333$
and $E_{\times 100} = 1075$ for \icfsouth. As in our primary
analysis, the highest-excursion events in the scrambled data are due
to the two GRBs observed in the \icfnorth\ data. Excluding these GRBs
would give excursion thresholds of $E_{\times 10} =575$ and
$E_{\times 100} = 1102$ for \icfnorth.

Analyzing the unscrambled IC59 datasets reveals no excursions above
the $E_{\times 10}$ threshold for either dataset. Performing the
Anderson-Darling test on the null and unscrambled distributions yields
$p=55\%$ for \icfnorth\ and $p=62\%$ for \icfsouth. We therefore see
no evidence for spatio-temporal correlation of the neutrinos and
\fermi\ gamma-rays on the $\sim$0.5~day timescale that we probed.



We conclude that the observed tendency of IC59 neutrinos to arrive
from brighter portions of the \fermi\ gamma-ray sky, while potentially
due to a statistical fluctuation (single-trial $p=7.0\%$ for the two
hemispheres combined), is both intriguing in its own right and likely
explains the systematic trends in $\lambda$ residuals against
scrambled datasets observed for both hemispheres in our original
analysis (single-trial $p=3.9\%$ for the two hemispheres
combined). While this $p$-value cannot support an evidence or
discovery claim in the context of our multistage analysis, it
nonetheless points the way to interesting future analyses that could
make use of eight further years of \ice\ data from the 79-string and
full-strength (86-string) arrays.

In particular, we note that it is a low-level (single neutrino)
correlation between the neutrino and gamma-ray skies that has prompted
current interest in the blazar \txsblazar\ and its possible neutrino
\mbox{IceCube-170922A} \citep{1709blazaratel}.  


\section{Conclusions}
\label{sec:conc}

We have carried out an archival coincidence search for neutrino +
gamma-ray emitting transients using publicly available \fermi\ LAT
gamma-ray data and \ice\ neutrino data from its 40-string and
59-string runs, incorporating \fermi\ data from the start of mission
in Aug~2008 through May~2010. Our search was designed to be capable of
identifying \nugamma\ transients either as individual
high-significance single-neutrino events with high gamma-ray
multiplicity, or as a population, via statistical comparison of the
observed pseudo-likelihood distributions to those of uncorrelated
(scrambled) datasets.

Using Monte Carlo simulations and signal injection, we demonstrated
sensitivity to single-neutrino events of sufficient gamma-ray
multiplicity. High-multiplicity gamma-ray clusters have been observed
throughout the \fermi\ mission in coincidence with bright LAT-detected
gamma-ray bursts, including two bursts occurring during our period of
study, GRB\,090902B ($>$200 photons; \citealt{latgrb090902}) and
GRB\,100414A ($>$20 photons; \citealt{latgrb100414}).

We established sensitivity to subthreshold populations of transient
\nugamma\ sources at the $>$13\% (IC40), $>$9\% (\icfnorth), and
$>$8\% (\icfsouth) level for the three hemisphere-specific neutrino
datasets we analyzed ($p=1\%$ threshold;
Sec.~\ref{sub:scrambled}). These limits are expressed as the fraction
of all neutrinos present in the datasets that are due to
\nugamma\ transient sources ($\delta t < 100$\,s), according to our
assumptions (Sec.~\ref{sub:injection}). Expressed as event rates, the
limits correspond to $>$210 (IC40), $>$440 (\icfnorth), and $>$565
(\icfsouth) gamma-ray associated neutrinos per hemisphere per
year. Sensitivity of a joint analysis of the IC59 datasets was not
separately established but can be estimated at $>$850 gamma-ray
associated neutrinos per year all-sky. While these limits are well
above the conservative limit of $r_{\rm cosmic} \simgt 120$~neutrinos
per year all-sky for the full detector array that we derive on the
basis of the $\enu \simgt 60$\,TeV cosmic neutrino spectrum
(Sec.~\ref{sec:intro}), that rate could be substantially larger if the
cosmic neutrino spectrum softens significantly within the ${\rm
  1\,TeV} \simlt \enu \simlt {\rm 60\,TeV}$ range relevant to these
data.

Unscrambling the neutrino data, we identify no individual
high-significance neutrino + high gamma-multiplicity events, and no
significant deviations from the null test statistic ($\lambda$)
distributions. However, we observe a similar and unexpected pattern in
the $\lambda$ residuals from the \icfnorth\ and \icfsouth\ analyses,
our two more sensitive datasets, corresponding to a joint $p$-value of
3.9\% (Sec.~\ref{sub:unscrambled}). While granting that these residual
patterns may be due to statistical fluctuations, we carried out
additional investigations in an attempt to determine the origin of the
deviations, and whether or not they suggest the presence of
\nugamma\ correlated emission. 

We first vetted the IC59 data for short timescale transients (our
original test hypothesis) in two ways, checking for systematic trends
in the temporal and spatial separations of the neutrino event and its
associated gamma-rays. No systematic trends in spatial or temporal
separation are evident for either \icfnorth\ or
\icfsouth\ (Sec.~\ref{sub:vetting}).

We then checked for \nugamma\ spatio-temporal correlations on
timescales beyond our original $\pm$100~s window
(Sec.~\ref{sub:correlations}). We searched for neutrino-correlated
gamma-ray flux excursions within a $\pm$50,000~s ($\sim$0.5~day)
window centered on the neutrino arrival time, finding no evidence for
correlated gamma-ray flux excursions on this timescale.  Instead, we
find a likely correlation ($p=7.0\%$, single trial) of IC59 neutrino
positions with persistently bright portions of the \fermi\ gamma-ray
sky.

This interesting and unexpected finding of our search for cosmic
\nugamma\ sources, a possible signature of gamma-ray correlated
structure in the high-energy neutrino sky, should be readily testable
using eight years of further data already collected by the 79-string
and full-strength (86-string) \ice. 

In particular, if blazars are responsible for a non-negligible
fraction of the highest-energy cosmic neutrinos, then -- given the
brightness of the blazar population over the ${\rm 100\,MeV} \simlt
\egam \simlt {\rm 300\,GeV}$ LAT bandpass -- this would generate
correlated structure in lower-energy neutrinos. Blazar associations
have been proposed for two likely-cosmic high-energy neutrinos,
\mbox{IceCube-121204} ``Big Bird'' and \mbox{IceCube-170922A}, thanks
to their spatio-temporal proximity to flaring episodes of the blazars
PKS~B1424$-$418 \citep{kkm+16} and \txsblazar\ \citep{1709blazaratel},
respectively. On the other hand, blazar models are strongly
constrained by the \ice\ \fermi-blazar stacking anlaysis
\citep{ice17blazars}, and by the absence of detected neutrino point
sources \citep{ice15ptsrc,murasewaxman16}.

In a general sense, some level of correlation between the gamma-ray
and neutrino skies is anticipated in models that propose a common
origin for the diffuse $\enu\simgt 100$\,TeV neutrino and $\egam\simlt
1$\,TeV gamma-ray backgrounds \citep{mal13,fangmurase18}.

Finally, production of some cosmic neutrinos by Galactic sources,
whether compact binaries \citep{milagro12cygnus,AGPestimate2014}, TeV
unidentified sources or hypernova remnants
\citep{Budnik+08hn,fkm13,muraseahlers14}, or other source
population(s), would naturally lead to correlated structure, given the
very prominent Galactic signature in \fermi\ all-sky maps
(Fig.~\ref{fig:phbkg}).

Looking ahead, we eagerly anticipate the results of a systematic and
comprehensive search for \fermi\ gamma-ray correlated structure in the
full \ice\ dataset. In addition, having demonstrated its effectiveness
on archival data, we will be working with \ice, ANTARES
\citep{antaresdesg}, and other partner facilities of the Astrophysical
Multimessenger Observatory Network (AMON) to deploy our neutrino +
high gamma-multiplicity search and generate low-latency (delays of
$\approx$5~hours) \nugamma\ alerts from \fermi\ LAT gamma-ray and
\ice\ and \antares\ neutrino data. These AMON alerts will be
distributed in real-time to AMON follow-up partners, prompting
rapid-response follow-up observations across the electromagnetic
spectrum.



\acknowledgments

The authors thank Erik Blaufuss and David Thompson for helpful
discussions. We gratefully acknowledge support from Penn State's
Office of the Senior Vice President for Research, the Eberly College
of Science, and the Penn State Institute for Gravitation and the
Cosmos. This work was supported in part by the National Science
Foundation under Grant Number \mbox{PHY-1708146}. K.~M. is supported
by the Alfred P. Sloan Foundation and by the National Science
Foundation under Grant Number \mbox{PHY-1620777}.

\software{Astropy \citep{astropy}, Matplotlib \citep{matplotlib},
  HEASoft \citep{heasoft}, HEALPix \citep{healpix}, SciPy \citep{scipy}}



\bibliographystyle{aasjournal}
\bibliography{fic}


\end{document}